\documentclass[showpacs,superscriptaddress, preprintnumbers,amsmath,amssymb,twocolumn]{revtex4}

\usepackage{graphicx}
\usepackage{dcolumn}
\usepackage{bm}
 
\usepackage{feynmf}
\usepackage{times}
\include{epsf}

\begin{document}

\unitlength = 1mm

\title{

Finite size effects. The averaged eigenvalue density of  Wigner random sign  real symmetric  matrices.
}   
\author{  G.S.Dhesi }
\affiliation{  
London South Bank University, SE1 0AA, United Kingdom\\
Email : dhesig@lsbu.ac.uk }
 \author{ M. Ausloos}
\affiliation{  
 GRAPES, rue de la belle jardini\`ere 483/002, B-4031 Li\`ege Angleur Sart-Tilman, Belgium, Europe\\ email: marcel.ausloos@ulg.ac.be
 \\ and \\
 School of Management,
University of Leicester,  University Road, Leicester, LE1 7RH, UK  \\  $email$: ma683@le.ac.uk }
   
\preprint{revised version for PRE    }

  \date{\today}
 
\begin{abstract}
 Nowadays, strict finite size effects must be taken into account  in condensed matter problems when treated through models based on lattices or graphs. On the other hand, the  cases of directed bonds or links are known as highly relevant, in topics ranging from ferroelectrics to quotation networks. Combining these two points leads to examine finite size random matrices.  To obtain basic materials properties, the  Green function  associated to the matrix has to be calculated.  In order to obtain the first  finite size  correction   
a perturbative scheme is hereby developed  within the framework of the replica method.  The averaged eigenvalue spectrum and the corresponding Green function of Wigner random sign real symmetric  $N$ x $N$ matrices to order $1/N$ are {\it in fine} obtained analytically.  Related simulation  results are also presented. The comparison between the analytical formulae and finite size matrices numerical diagonalization results exhibits an excellent agreement, confirming the correctness of  the first order finite size expression.
\end{abstract}



 \maketitle
 



\section{Introduction} \label{sec1} 

 Random matrix studies can be traced back a long time ago, but are an intense research subject nowadays \cite{Tao}. They were an essential part of discoveries in nuclear physics  \cite{Wigner1,Wigner2}.  Thereafter,  the emergence of amorphous and/or  disordered alloys as interesting materials led to consider the distribution of eigenvalues, $\lambda$,  for  the Green's function  (itself a random matrix) associated to the Hamiltonian describing  the system \cite{JNCS810FA}; see Eq.(\ref{eq2.8}) below.

 In brief, it is well known  that the density of states is the imaginary  ($Im$) part of this Green's function, the energy states being the eigenvalues  \cite{MehtaGaudin,KozmaKorniss}.  Thus, the eigenvalue spectrum has to be well known,  in particular  in order to determine the presence of  (optical or conduction)  spectral gaps and state  localization,  like in the context of Anderson  model  \cite{Slalina2006,shklovskii1993statistics} and spin glass models \cite{parisi1993spinglassl}. Moreover, the largest  (necessarily real according to the Perron-Frobenius) and the next to largest eigenvalues  (not necessarily real in the case of asymmetric matrices) are, for the former, indications of the ground state of the system, through  an approximation of the free energy and of the diffusion  (or relaxation) coefficient, for the latter.  
 From a  "more general" point of view, let us simply say that the  Averaged Eigenvalue Density (AED) and its properties have to be  calculated, or must receive some theoretical  estimate with enough precision taking into account the system finite size,   - the best being when searching for universal features.

 There is a large body of mathematical work on the spectra of   
  random  matrices, ranging from modern versions of Perron-Frobenius theorem for non-negative matrices \cite{Perron,Frobenius},  - up to recent results reviewed by  Brualdi   \cite{Brualdi}  and others like in  \cite{Bapat and Raghavan 1997,Berman Neumann and Stern 1989,Berman and Plemmons 1994,Minc 1988, Rothblum 2006,Senata 1981}.    In fact,  a  paper on  "finite size corrections to disordered Ising models", on random regular graphs \cite{dRRG},   recently appeared  \cite{PRE90.14.012146},   indicating the up-to-date  interest  of considering binary distributions in the context of random matrices and graphs.  When this paper was in its final stage, a review \cite{RME87.15.1037} appeared on the "random-matrix theory of Majorana fermions and topological superconductors", indicating the up-to-date  interest in the matter. 
  
One of the most studied ensembles of random matrices is the Gaussian Orthogonal Ensemble (GOE), of $N$ x $N$ real symmetric matrices which is invariant under orthogonal transformations   \cite{JPhAMT40.07.1561,delannay2000distribution,JPhAMT42.09.035001,JPhAMT46.13.305203}.   
Much attention has been paid to the calculation of its  AED. It was shown by Wigner \cite{W58} that the AED for the GOE in the limiting case of $N$  going to infinity is a semicircle. This, amongst many other early results in the field of random matrices, can be found in the early  texts by Porter \cite{Porter 1965} and by Mehta \cite{Mehta 1967}. {  Most of the investigation methods   rely either on elaborate moment and cumulant expansions or on the properties of Orthogonal Polynomials \cite{Pastur,Deift,Konig}.  A radically different method was presented by Edwards and Jones \cite{EJ76} for calculating the AED. The  Edwards and Jones (EJ) method relied on the so called "replica trick", first employed by Edwards  \cite{E70} in the study of polymer physics, reviewed by   Advani et al., \cite{Advani et al 2013} for example, which led to the renormalization 
group technique later on.

 By no means,  all the published work on random matrices has  been directed at the GOE. Wigner   \cite{W55} addressed the problem of calculating the AED of an ensemble of large symmetric random matrices that were either bordered, i.e. having   integers along  the diagonal and random numbers equal to plus or minus some constant $J$ on  the super- and sub-diagonals, or  had zeros on the diagonal and entries that (subject to the symmetry requirement) were either  $+J$ or  $-J$ in the off-diagonal elements. For example, let us have in mind an Ising spin system in which two neighbors are pointing in different or in similar directions,  thus   resulting in  a $-J$ or $+J$   bonding energy;  dipoles in   ferroelectric materials can also be considered as being at different energy levels antisymmetrically placed with respect to the 0 level. Another case is the fully connected network where links can take two different  weights, - here  they should be  equal in magnitude but with opposite signs.
 The latter of these two ensembles is  best to be called the random sign symmetric matrix ensemble (RSSME). 
 
  In a short article, Wigner  \cite{W58} conjectured that a semicircular distribution of eigenvalues would be the limiting distribution obtained as  $N \rightarrow \infty$ for an ensemble of symmetric matrices in which the probability density function (pdf) of any off-diagonal element is reasonably well behaved and  for  which the second moment of all such off-diagonal elements should have the same constant value.  It is worth pointing out that this is akin to the AED  of a $d-$regular random graph, with $N$ vertices, for $N \rightarrow \infty$ and  $d \rightarrow \infty$. This is due to the tree-like structures that emerge when  calculating the AED of these sets of random matrices.
   Hence,  the AED   becomes a semicircle as in the GOE of matrices, 
     in  the limit of  size $N \rightarrow \infty$,   and 
diverging mean 
  $d \rightarrow \infty$, 
i.e. when the Kesten-Mckay law converges to a semi-circular law  \cite{KMlawKesten,KMlawMcKay,KMLawjakobson1999eigenvalue,KMLawneri2012spectra,KMLawneritran2013sparse,Bauerschmidt} 

    However, no real system has an infinite size. The more so in  "real world" and "subsequent"   applications. The finite size constraint  must be nowadays  taken at its full value, even though it was previously often taken as an irrelevant non universal effect   in many condensed matter investigations. Very often, the surface energy  and surface entropy terms were disregarded in free energy calculations.

 Jones and Dhesi  \cite{JD} 
 (hereafter referred to as JD)  applied the replica method to the case of  the RSSME and  showed, in the limit of  $N \rightarrow \infty$,  how easily the replica formalism produces a semicircular  AED. (Using the same formalism,  they were able to verify the  Wigner  conjecture.)
  This leads to the interesting problem of calculating the AED when $N$ is  finite. In  such a case ($N$ finite),  the AED   distribution departs from the  semicircle and becomes ensemble specific.   For the GOE, there has been a number of papers  that have addressed the problem of calculating the corrections to the Wigner semicircle which are of  order  $1/N$ \cite{TT,VZ,VWZ}.
  Dhesi and Jones 
 \cite{DJ90}  (thereafter referred  to as DJ) have provided a  comprehensive set   of results: DJ  calculated the AED for the GOE to order 1/$N^2$, and also the AED for finite $N$ based on a self-consistency argument, when each element of the matrix is drawn randomly  from a normal distribution.  Note that these four papers rely on different methods; furthermore,  the $1/N$ correction of Takano and Takano  \cite{TT}  differs from the  others.  
 
 Recently, Metz et al.   \cite{Metz Parisi Leuzi 2014} reconsidered finite size corrections to the spectrum of regular random graphs obtaining an analytical solution, given by a sum over loops comprising all length scales, each loop contributing with a term proportional to the difference of its effective resolvent with respect to the resolvent of an infinite closed chain.  In this context, let us also point to  Kanzieper  and  Akemann \cite{kanzieper2005statistics}  who looked "through the prism of  probabilities" on how  to find exactly  a given number of $real$ eigenvalues in the spectrum of an $N$x$N$ real  asymmetric Gaussian random matrix (see some elaboration on the conclusion section). {\it In fine}, to obtain an overview of relevant applications  and subsequent  approaches to complex systems, as those of concern here, see   \cite{KwapienDrozdz}. 

The present paper is still devoted to the calculation of the AED for the RSSME, { with vanishing diagonal elements, - though it will be shown that this  constraint is rather irrelevant,  to order $1/N$, but on an apparently more relevant set of cases,  like fully connected random graphs,  with a given distribution of  different types of links, see below.   Nevertheless, to provide a flavor of the   problem generality, let us compare the AED for the GOE  and the RSSME in both extreme cases, i.e.,  for $N$=2 or  $N \rightarrow \infty$.  When $N \rightarrow \infty$, the AEDs for both the GOE  and the RSSME obey the semicircular  distribution. However,  when $N$ = 2,  the  AED for the GOE   has the following form  \cite{PR} 

 \begin{equation} \label{eq1.1}
 \rho_{N=2}(\lambda) = \frac{1}{\sqrt{2\pi}}\;e^{-\lambda^2}\; [e^{-\lambda^2}+ \sqrt{\pi}\lambda\;erf(\lambda)]
  \end{equation}
where $erf$ is the error function \cite{Abramowitz}.

  For our ensemble,   when the diagonal elements  are 0,   the AED for the $N =2$  RSSME is merely given by two symmetric delta functions, per Eq.(\ref{eq3.1}), see  below in fact; see 
 Fig. \ref{fig: SimFig1NxN2x2} for  illustration.   Whence it can be noted that the departure from a semicircle, when $N$ is small, is much more acute in the case of the RSSME. Therefore,    the analytic result for the AED and the RSSME should describe the broadening and the overlapping of the peaked functions, thereby approaching  the semicircular function as $N$ becomes large.

  As it can be rather easily appreciated, this RSSME, i.e.  when an element is drawn randomly with equal probability of being positive or negative (head or tail),   is a more difficult exercise than when calculating  the case of the GOE,  i.e. when a normal type distribution (described by  Eq.(\ref{eq1.1}) as treated in DJ  \cite{DJ90}) converges to the semicircle as $N$ becomes large.  We will show that additional terms appear when the finite size is taken into account.

Therefore, the plan of this paper is the following. In Section \ref{sec2}, we  recall the basic replica technique for calculating the AED, and its corresponding Green function,  but geared toward the case of   random sign symmetric matrix ensembles. Section \ref{sec3} is devoted to casting the AED to order $1/N$ of the RSSME as a zero dimensional path integral. In Section  \ref{sec4}, we set up a perturbation theory, using Hubbard-Stratonovich transformation (auxiliary field identity) and Feynman diagrams,  which allows us  to calculate the correction to order $1/N$,  in Section  \ref{sec5}  with the steepest descent method. The correction is found to be non-vanishing and convergent only inside the Wigner semicircle and away from the band edges.   Analytic and simulation works are presented. The comparison between the analytical formulae and finite size matrices numerical diagonalisation results exhibits an excellent agreement, confirming the correctness of  the first order finite size expression. A few comments are made in Section \ref{comments}. 
 Numerical simulations are found in Sect. \ref{sec6NumericalSimulation}; their average is graphically compared to the theoretical expressions.

Finally, in Section  \ref{sec7conclusions}, we   summarize the results and suggest some direction   for further work, in view of possibly obtaining  related  results for more complicated cases and applications.
 
 \begin{figure}
 \includegraphics[height=6.0cm,width=8.4cm]{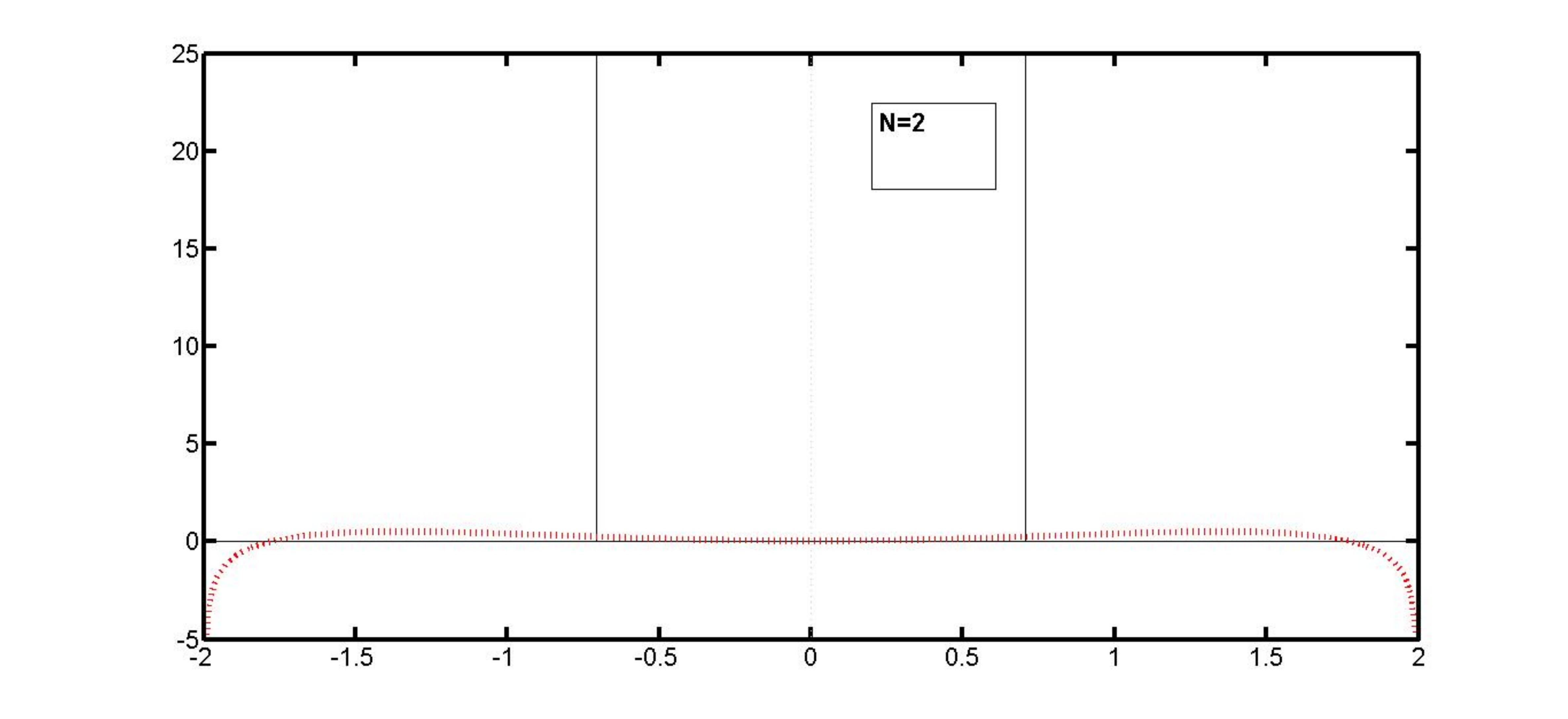}
  \caption   {
  Comparison of the $N =2$   "numerical simulation" AED  (vertical lines), leading   to 2 delta functions,  with the theoretical first order $O(1/N)$ approximation (red) dotted line,  resulting from Eq.(\ref{eq5.8}) and Eq.(\ref{eq5.11}). }
 \label{fig: SimFig1NxN2x2}
\end{figure}

\section{Replica technique} \label{sec2}
 
The theoretical development is based on the replica technique   \cite{E70}  which is briefly recalled for completeness within the present framework. Consider a real symmetric $N$ x $N$ matrix $J$ with eigenvalues $J_i$. 
The density  $v(\lambda)$ of such eigenvalues  is given by 
 \begin{equation} \label{eq2.1}
v(\lambda) = \frac{1}{ N}\; \sum_i \delta(\lambda-J_i)
  \end{equation}
 
where  $v(\lambda)$ has been chosen  to be normalized to unity.  For a real symmetric matrix, it can be  recalled  that 

 \begin{equation} \label{eq2.1det}
det (\Lambda -J) =  \prod_i (\lambda-J_i)
  \end{equation}
 where $\Lambda $ is the diagonal matrix with element  $\lambda$. In the complex plane, giving an infinitesimal imaginary part to $\lambda \rightarrow \lambda-i\epsilon$, one has
  \begin{equation} \label{eq2.2}
v(\lambda) = \frac{1}{N\pi}\; Im \frac{\partial}{\partial\lambda} ln\; det (\Lambda-J ).
  \end{equation}
  
  The replica trick,  further developed by Edwards and Jones   \cite{EJ76} in the context of random graphs,  uses
  
   \begin{equation} \label{eq2.3}
ln (x) =   \lim_{n\rightarrow 0} \Big[\frac{x^n-1}{n}\Big]
  \end{equation}
  so that Eq.(\ref{eq2.2}) reads
     \begin{equation} \label{eq2.4}
     v(\lambda) = \frac{-2}{ N\pi}\; Im  \frac{\partial}{\partial\lambda} \lim_{n\rightarrow 0} \frac{1}{n}
    \Big[\; det ^{-1/2}(\Lambda-J )^n-1\Big].
       \end{equation}
       
       The determinant ($det$)  can be parametrized as a multiple Fresnel integral \cite{Abramowitz,PRE87.13.012803rdmgraph}

        \begin{equation} \label{eq2.5}
    det ^{-1/2}(\Lambda-J ) =\Big(\frac{e^{i\pi/4}}{\pi^{1/2}}\Big)^N\;\int_{-\infty}^{\infty}\prod_idx_i e^{-i\sum_{i,j}x_i(\Lambda-J)_{i,j}x_j}.
       \end{equation}
       Substituting Eq.(\ref{eq2.5}) into Eq.(\ref{eq2.4}), assuming that this latter result holds for integer $n$ values, and can be continued for $n =0$, one obtains the fundamental result
        \begin{equation}    \label{eq2.6}
          \begin{array} {l}
  v(\lambda)  =  \frac{-2}{ N\pi}\; Im  \frac{\partial}{\partial\lambda} \lim_{n\rightarrow 0} \frac{1}{n} \Big \{\Big(\frac{e^{i\pi/4}}{\pi^{1/2}}\Big)^{Nn}\;\\\;\;\; \;\;\;\;\int_{-\infty}^{\infty}\prod_{i;\alpha}dx_{i;\alpha} [ exp(-i\Sigma_{i,j;\alpha}x_i^{\alpha}(\Lambda-J)_{i,j}x_j^{\alpha})-1]\Big \}
             \end{array}    \end{equation}
             
             The integration is now over the N$n$ variables $x_i^{\alpha}$ with $i\in(1,N)$ and $\alpha \in(1,n)$ respectively; the $ \lim_{n\rightarrow 0}$ being taken at the end of the calculation.    Therefore, the AED, $\rho(\lambda)$, of an ensemble of  real symmetric matrices 
             which has a  given pdf,  $p(J_{i,j})$,  is  
             
                  \begin{equation} \label{eq2.7}
\rho(\lambda)\equiv    < v(\lambda)> =  \int v(\lambda; J_{i,j}) \prod_{i,j} p(J_{i,j}) dJ_{i,j} 
       \end{equation}
 where the brackets  $<$ $>$ imply ensemble averaging. 
       
       Recall, at this stage, that the eigenvalue density is related to the Green function $\mathcal{G}(\lambda)\equiv(\Lambda-J ) ^{-1}$,  through
        \begin{equation} \label{eq2.8}
        v(\lambda)=    \frac{1}{\pi} Im  \frac{1}{N}  Tr \;\mathcal{G}(\lambda-i\epsilon )\end{equation} where $Tr$ stands for the trace 	and $\epsilon $ is supposed to be taken as small and positive; let us call $G(\lambda)$ the average Green function.
        Whence  the AED, $< v(\lambda)> \equiv  (1/\pi)\; Im\; G(\lambda)$, is   immediately obtained   from Eq.(\ref{eq2.8}) through the ensemble averaging.
       
       \section{Random sign symmetric matrix ensemble } \label{sec3} 
       
       Consider the specific case in which   a real symmetric matrix    (thus imposing $J_{i,j}=J_{j,i}$) has zero on its diagonal   ($J_{i,i}=0$)}, but the off-diagonal elements  $J_{i,j}$ take randomly the value $+J/\sqrt N$ or  $-J/\sqrt N$ with equal probability 0.5.  Practically,  the sign can indicate whether a bond or link is directed or not, pertains to a ferromagnetic or antiferromagnetic set of spins, or has a given color, for example; the equal probability constraint  and the matrix symmetry will be suggested, in SectIon  \ref{sec7conclusions},  to be removed in further work. Let $J$ be of the order of unity. The ensemble pdf is described by 
        \begin{equation} \label{eq3.1}
    p(J_{i,j})    = \frac{1}{2}\{\delta (J_{i,j}-J/\sqrt N )+\delta (J_{i,j}+J/\sqrt N )
    \}.\end{equation}
    Substituting Eq.(\ref{eq3.1}) into Eq.(\ref{eq2.6}) and Eq.(\ref{eq2.7}), the integral over $J_{i,j}$ is next performed.
    
     \begin{equation}   
          \begin{array} {l}
  \rho(\lambda)  =-\frac{2}{N\pi} Im \frac{\partial}{\partial\lambda}  \lim_{n\rightarrow 0} \frac{1}{n}  \Big \{\Big(\frac{e^{i\pi/4}}{\pi^{1/2}}\Big)^{Nn} \; \\ \\\ \ \ \ \ \ \int_{-\infty}^{\infty}\prod_{i;\alpha}dx_{i;\alpha} \Big[ exp(-i\;\lambda \sum_{i;\alpha}(x_i^{\alpha})^2\Big]\; \\ \\\ \ \ \ \ \ \ \;\prod_{i<j}\Big[ cos(  \frac{2J}{\sqrt N}  \sum_{\alpha}x_i^{\alpha}x_j^{\alpha})-1\Big]\Big\}
             \end{array}    \end{equation} \label{eq3.2}
             Of course, 
               \begin{equation}   
               \prod_{i<j}cos\Big(  \frac{2J}{\sqrt N}  \sum_{\alpha}x_i^{\alpha}x_j^{\alpha}\Big) \equiv exp\Big\{\frac{1}{2}\sum_{i,j} ln  \Big[cos(  \frac{2J}{\sqrt N}  \sum_{\alpha}x_i^{\alpha}x_j^{\alpha})\Big]\Big\}.
                 \end{equation} \label{eq3.3}
               It should be pointed out,  thanks to a comment by a reviewer,  that   the diagonal elements of the rhs in this transformation,   should be more thoroughly discussed;  see Appendix. Thereafter the argument of the exponential can be expanded in powers of 1/N to read
                 
                     \begin{equation}   \label{eq3.4}
          \begin{array} {l}
                   \prod_{i<j}cos\Big(  \frac{2J}{\sqrt N}  \sum_{\alpha}x_i^{\alpha}x_j^{\alpha}\Big) \simeq
                   \;\\\  \\\ \ \ \   exp\Big\{ \sum_{i,j}\Big (  \frac{-J^2}{ N} ( \sum_{\alpha}x_i^{\alpha}x_j^{\alpha})^2\Big) \Big [1+\frac{2J^2}{3N}  ( \sum_{\alpha}x_i^{\alpha}x_j^{\alpha})^2\Big] \; \\ \\\ \ \ \ \    \ \ \ \    \ \ \ \  \ \ \ \    \ \ \ \  \ \ \ \    \ \ \ \    \ \ \ \  \ \ \ \    \ \ \ \     + O(N^{-3})\Big\}
             \end{array}    \end{equation} 
             
      Observe the term in brackets in Eq.(\ref{eq3.4}). It will lead to the relevant term in Eq.(\ref{eq4.4}) distinguishing  a difference
between
matrices
with
Gaussian
distributed
matrix
elements
and
those
with
the
binary
distribution
considered
here, and subsequently to Eq.(\ref {eq4.15}  ) and Eq.(\ref{eqrhohs})  for  the  (1/N) correction to $\rho(\lambda)$.

             Keeping only the first term of the exponential and following the JD analysis, the AED is easily obtained  in the limit  $N\rightarrow \infty$, i.e. a semicircle. In order to obtain the finite size $N$ case,  the   next  leading term must  be conserved.
             
                    \begin{equation}   
          \begin{array} {l}
  \frac{-J^2}{ N}     \sum_{i,j}  ( \sum_{\alpha}x_i^{\alpha}x_j^{\alpha})^2   =   \frac{-J^2}{ N}     \sum_{\alpha} [ \sum_{i}(x_i^{\alpha} )^2 ]^2 +   \;\\\ \\ \ \ \ \    \ \ \ \    \ \ \ \  \ \ \ \    \ \ \ \  \ \ \ \    \frac{-J^2}{ N}     \sum_{i,j}  ( \sum_{\alpha \neq  \beta}x_i^{\alpha}x_j^{\alpha} x_i^{\beta}x_j^{\beta} ).
           \end{array}    \end{equation} \label{eq3.5}
           
           As in Edwards and Warner  \cite{EW80}  and Dhesi and Jones  \cite{DJ90}, it can be shown that the first term is an order of magnitude higher than the second when $N\rightarrow \infty$.  Indeed, even though each $\alpha \neq \beta$ terms tend to contribute to the AED, the sum over roman indices $i$ and $j$ reduces their input.
           
           Nevertheless when calculating the next contribution to the AED the $\alpha \neq \beta$ terms  must be conserved. However, the second term  can be decomposed into $\alpha, \beta, \gamma, \delta$ contributions. Again, the $\alpha=\beta= \gamma= \delta$ terms will contribute to an order of magnitude larger value than those with non equal indices. The full formal expression is not written for conciseness; see below for its practical evaluation, Eq.(\ref{eq4.3}).
           
             \section{Perturbation method} \label{sec4} 
             
             In this section, the starting idea is to use a Hubbard-Stratonovich transformation (or  so called auxiliary field identity) so as to express the second and third term in the exponential
             
             \begin{equation}   
          \begin{array} {l} \label{eq4.1}
          exp\Big\{ \frac{-J^2}{N} [ \sum_{i}(x_i^{\alpha} )^2 ]^2\Big\}=
          \frac{N^{1/2}}{(2\pi)^{1/2}}    \frac{1}{(2J^2)^{1/2}}  \lambda
           \;\\ \\\ \ \ \ \    \ \ \ \    \ \ \ \  \ \ \ \    \ \ \ \  \ \ \ \     \ \ \ \  \ \ \ \     
           \int_{-\infty}^{\infty} ds^{\alpha} \;exp\Big[-\frac{\lambda^2}{4J^2}N(s^\alpha)^2                       \;\\\ \\ \ \ \ \    \ \ \ \    \ \ \ \  \ \ \ \    \ \ \ \  \ \ \ \    \ \ \ \  \ \ \ \   
                      -i\lambda s^\alpha\sum_i (x_i^\alpha)^2\Big]
            \end{array}    \end{equation} 
            and similarly for
              \begin{equation}   
          \begin{array} {l}
          exp \Big\{ \frac{-2J^4}{3N^2} [ \sum_{i}(x_i^{\alpha} )^4 ]^2\Big\} = \sqrt{\frac{3}{8\pi} } \int_{-\infty}^{\infty} dp^{\alpha} \;exp\Big[ -\frac{3}{8}(p^\alpha)^2 \;\\\ \\ \ \ \ \    \ \ \ \  \ \ \ \  \ \ \ \      \ \ \ \  \ \ \ \    \ \ \ \  \ \ \ \    \ \ \ \  \ \ \ \   -i \frac{J^2}{N}p^{\alpha}\sum_i(x_i^\alpha)^4\Big]
            \end{array}    \end{equation} \label{eq4.2}
            
            Gathering all the relevant terms,  the AED reads
             \begin{equation}   
          \begin{array} {l}  \label{eq4.3} 
     \rho(\lambda)=-\frac{-2}{N\pi}  \frac{\partial}{\partial\lambda}  \lim_{n\rightarrow 0} \frac{1}{n}
    \Big \{\Big(\frac{e^{i\pi/4}}{\pi^{1/2}}\Big)^{Nn} \\\\\ \ \ \ \ \ \int \prod_{\alpha}ds_{\alpha}
  
  \frac{N^{1/2}}{(2\pi)^{1/2}} \frac{1}{(2J^2)^{1/2}} \lambda  \;exp\Big[-\frac{\lambda^2}{4J^2} N\sum_\alpha(s^\alpha)^2\Big]
  \\\\\ \ \ \ \ \ \int \prod_{\alpha}dp_{\alpha} \sqrt{\frac{3}{8\pi}} exp\Big[-\frac{3}{8}\sum_\alpha (p^\alpha)^2\Big]
      \;\;  L(s;p)-1
    \Big   \}\; 
           ,  \end{array}    \end{equation}
            
            where 
            
             \begin{equation}   
          \begin{array} {l} \label{eq4.4} 
          L(s;p)= \int \prod_{i;\alpha}dx_{i;\alpha}
exp  \Big   \{
-i\lambda \sum_{i;\alpha}  (x_i^{\alpha} )^2    
 \\\\\ \ \ \ \     \ \ \ \    \ \ \ \  \ \ \ \    \ \ \ \  \ \ \ \ 
-i\lambda s^\alpha \sum_{i;\alpha}  (x_i^{\alpha} )^2  
   \;\\\\\ \ \ \ \    \ \ \ \    \ \ \ \  \ \ \ \    \ \ \ \  \ \ \ \   -i \frac{ J^2}{ N} p^\alpha   \sum_{i;\alpha}  (x_i^{\alpha})^4 
 \;\\\\\ \ \ \ \    \ \ \ \    \ \ \ \  \ \ \ \    \ \ \ \  \ \ \ \  
 -   \frac{ J^2}{ N}( \sum_{i,j;\alpha \neq  \beta}x_i^{\alpha}x_j^{\alpha} x_i^{\beta}x_j^{\beta} ) 
  \Big   \}\; 
              \end{array}    \end{equation} 
              
              From a close examination of Eq.(\ref{eq4.4}), it can be noticed that the third and fourth terms in the exponential do $not$ contribute to the AED  in the $ \lim  N\rightarrow \infty$. Thus a "perturbation expansion" can  be constructed in order to represent $L(s;p) $ as 
                 \begin{equation}   
          \begin{array} {l} \label{eq4.5} 
          L(s;p)= \int \prod_{i;\alpha}dx_i^{\alpha}\;
         \Big[ exp\big(-i\lambda(1+s^\alpha) \sum_{i;\alpha} (x_i^\alpha)^2\big)\Big] \\Ê\\  \ \ \ \    \ \ \ \  \ \ \ \    \ \ \ \  \ \ \ \  (1+A  ) \;(1+B )   .    \end{array}    \end{equation} 
          
          with
                \begin{equation}   
          \begin{array} {l} \label{eq4.6}
          A=   -i \frac{ J^2}{ N} p^\alpha   \sum_{i;\alpha}  (x_i^{\alpha})^4 
 \;\\\\\ \ \ \ \    \ \ \ \    \ \ \ \  \ \ \ \    
 + \frac{1}{2!} (-i\; \frac{ J^2}{ N} p^\alpha )^2  (\sum_{i;\alpha}  (x_i^{\alpha})^4 )^2+ \dots
          , \end{array}    \end{equation} 
          and
           \begin{equation}   
          \begin{array} {l} \label{eq4.7}
       B=       \frac{ -J^2}{ N}( \sum_{i,j;\alpha \neq  \beta}x_i^{\alpha}x_j^{\alpha} x_i^{\beta}x_j^{\beta} )   \;\\\\\ \ \ \ \    \ \ \ \    \ \ \ \  \ \ \ \    
 + \frac{1}{2!} (\frac{ -J^2}{ N})^2\;( \sum_{i,j;\alpha \neq  \beta}x_i^{\alpha}x_j^{\alpha} x_i^{\beta}x_j^{\beta})^2
  + \dots \end{array}    \end{equation} 
   
Next, consider the part of  $L$ involving the $A.B$ term, 
 denoting it  by $ L_{ AB   }$, i.e., 
\begin{equation} \begin{array} {l} \label{eq4.9}
 L_{ AB   }= \int \Pi_{i,\alpha} dx_i^{\alpha}  exp[-i\lambda(1+s^{\alpha} )\Sigma_{i;\alpha}  x_i^{\alpha^2}  ]\;A.B \;\\ \\  \ \ \ \    \ \ \ \    \ \ \ \  \ \ \ \    \equiv \Sigma_{c,d}\; L_{ AB cd }
       , \end{array}   \end{equation}
where $ L_{ AB cd }$ is the contribution to $ L_{ AB   }$ from the $c-th$ and $d-th$ terms of 
$A$ and $B$ (where $c$ and $d$ are positive integers). 
We employ a diagrammatic technique \cite{PRE90.14.010102R} to evaluate  $ L_{ AB}$ by 
representing 
\vskip 0.45cm


\begin{eqnarray} 
\sum_{k;\gamma} \;(x_k^\gamma)^4 & \rightarrow & \parbox{35mm} 
{ \begin{fmffile}{simple}
\begin{fmfgraph*}(20,10)
\fmfleft{i1,i2}
\fmflabel{$k \gamma$}{i1}
\fmflabel{$k \gamma$}{i2}
\fmfright{o1,o2}
\fmflabel{$k \gamma$}{o1}
\fmflabel{$k \gamma$}{o2}
\fmf{plain}{i1,o2}
\fmf{plain}{i2,o1}
\end{fmfgraph*}
\end{fmffile} }
\label{eq.1FD}
\end{eqnarray}

and 


\begin{eqnarray} 
\sum_{i,j;\alpha \neq  \beta}x_i^{\alpha}x_j^{\alpha} x_i^{\beta}x_j^{\beta}\; & \rightarrow &  \parbox{35mm}{ \begin{fmffile}{simple_labels}
\begin{fmfgraph*}(40,25)
\fmfleft{i1,i2}
\fmfright{o1,o2}
\fmf{plain,label=$i \alpha $}{i1,v1}
\fmf{plain,label=$i \beta $}{v1,i2}
\fmf{plain,label=$j \alpha $}{o1,v2}
\fmf{plain,label=$j \beta $}{v2,o2}
\fmf{dashes}{v1,v2}
\end{fmfgraph*}
\end{fmffile}}  \label{eq25}
\end{eqnarray}

 \begin{widetext}
Thus, taking   the first term from both A and B we can represent $L_{AB11}$ symbolically as

\vskip 0.5cm
\begin{eqnarray}  
 -i\frac{J^4}{N^2} \; p^\gamma\; & \Big\{ \;\; \parbox{40mm}{ \begin{fmffile}{simple2}
\begin{fmfgraph*}(30,20) \label{eq26}
\fmfleft{i1,i2}
\fmfright{o1,o2}
\fmf{plain}{i1,o2}
\fmf{plain}{i2,o1}
\fmflabel{$k \gamma$}{i1}
\fmflabel{$k \gamma$}{i2}
\fmflabel{$k \gamma$}{o1}
\fmflabel{$k \gamma$}{o2}
\end{fmfgraph*}
\end{fmffile} } & \parbox{40mm}{ \begin{fmffile}{simple_labels2}
\begin{fmfgraph*}(35,25)
\fmfleft{i1,i2}
\fmfright{o1,o2}
\fmf{plain,label=$i \alpha $}{i1,v1}
\fmf{plain,label=$i \beta $}{v1,i2}
\fmf{plain,label=$j \alpha $}{o1,v2}
\fmf{plain,label=$j \beta $}{v2,o2}
\fmf{dashes}{v1,v2}
\end{fmfgraph*}
\end{fmffile}
} \Big\} 
\end{eqnarray} 

\end{widetext}

where the brackets  $\Big\{$  $\Big\}$  denote the average against the 
Gaussian weight in Eq. (\ref{eq4.9}). 

Following the usual diagrammatic technique plus bearing in mind  that $\alpha \neq \beta$ we evaluate the integral
 defining  $L_{ AB   }$ by contracting  the legs. In so doing, we produce either connected or disconnects diagrams

\subsection{Disconnected Diagram}

The disconnected diagram reads
\vskip 0.5cm

\begin{eqnarray}  
& \begin{array}{c}
            k \gamma \\ k \gamma
           \end{array} \!\!\!\!\!\!\!\!\!\!\!\!\!\!\!\! \parbox{20mm}{ \begin{fmffile}{ears}
\begin{fmfgraph*}(25,5)
\fmfleft{i,i1}
\fmfright{o,o1}
\fmf{phantom}{i,v,v,o}
\fmf{plain}{v,v}
\fmf{plain,left=90}{v,v}
\end{fmfgraph*}
\end{fmffile} }\!\!\!\!\!\!\!\! \begin{array}{c}
            k \gamma \\ k \gamma
           \end{array} & \begin{array}{c}
            i \alpha \\ j \beta
           \end{array} \!\!\!\!\!\!\!\! \parbox{20mm}{ \begin{fmffile}{disconnected2}
\begin{fmfgraph*}(25,5)
\fmfleft{i,i1}
\fmfright{o,o1}
\fmf{phantom,tension=5}{i,v1}
\fmf{phantom,tension=5}{v2,o}
\fmf{plain,left,tension=0.4}{v1,v2,v1}
\fmf{dashes}{v1,v2}
\end{fmfgraph*}
\end{fmffile}
} \!\!\!\!\!\!\!\! \begin{array}{c}
           \;\; i \alpha \\ \;\; j \beta
           \end{array}
\end{eqnarray} 

\vskip 0.2cm
This diagram gives a contribution whose $n$-dependence is of the $O(n(n^2-n))$.    Remembering that we have to take the limit $n \rightarrow 0$, in evaluating the AED, it becomes clear (also see DJ \cite{DJ90}) that to produce a non zero contribution to the AED, we need to retain terms that are linear in $n$. Therefore the \underline{above} diagram contributes nothing to the AED.  It can be seen that a necessary condition for a diagram to be linear in $Nn $ is that it be connected. This result is general and holds to all orders in  perturbation theory \cite{PRB.13.76.1329}.

\subsection{Connected Diagram }
Bearing in mind  that $\alpha \neq \beta$,  the connected diagram that contributes of $O(1/N)$  to $L_{ AB;11   }$ is

\vskip 0.5cm

 \begin{equation} 
 \begin{fmffile}{connected}
\begin{fmfgraph*}(50,30)
\fmftop{t0,t1,t,t2,t3}
\fmfbottom{b0,b1,b,b2,b3}
\fmf{phantom}{t1,v1,b1}
\fmf{phantom}{t2,v2,b2}
\fmf{phantom,tension=40}{t,v3,v31,v32,v33,v34,v35,v36,v37,v38,v39,v4,b}
\fmf{dashes}{v3,v4}
\fmffreeze
\fmf{phantom}{v1,v0,t}
\fmffreeze
\fmf{phantom}{v1,va,b}
\fmf{phantom}{v1,vt,t0}
\fmf{phantom}{v1,vb,b0}
\fmf{plain,right}{v1,v2,v1}
\fmf{plain,tension=0.8,right=270}{v1,v1}
\fmflabel{$i \beta \;\;\; i \alpha$}{v3}
\fmflabel{$j \beta \;\;\; j \alpha$}{v4}
\fmflabel{$ k \gamma$}{v0}
\fmflabel{$ k \gamma$}{va}
\fmflabel{$ k \gamma$}{vt}
\fmflabel{$ k \gamma$}{vb}
\end{fmfgraph*}
\end{fmffile}
\end{equation}

\vskip 0.2cm
Since $\alpha \neq \beta$,  then  $i=j=k$. This in turn provides a contribution to AED of  $O(1/N^2)$ 
Furthermore, it
is realized that any general term of  $L_{ABcd}$  will give rise to connected diagrams that are linear in $ n$ will be to a maximum to $O(N^{-1})$. 

Subsequently  $L_{AB}$  contributes to the AED to a maximum of $O(N^{-2})$. As we are to evaluate the AED to $O(N^{-1})$, this  $L_{AB}$   part of $L$ can thus be disregarded. 

Whence Eq.(\ref{eq4.4}) can be rewritten as 
\begin{equation} \label{eq4.10}
L =    I\; (1+M + K)
\end{equation}
 where 
\begin{equation} \label{eqI}
I =  \int \prod_{i;\alpha} dx_{i;\alpha} \; exp\big[ -i\lambda (1+s^\alpha)\;\sum_{i;\alpha}(x_i^\alpha)^2\big].
\end{equation}
\begin{equation} \label{eqM}
M=    \int \prod_{i;\alpha} dx_{i;\alpha} \; exp\big[ -i \frac{J^2}{N}p^\alpha\;\sum_{i;\alpha}(x_i^\alpha)^4\big].
\end{equation}
\begin{equation} \label{eqK}
K=     \int \prod_{i;\alpha} dx_{i;\alpha} \; exp\big[ -i \frac{J^2}{N}p^\alpha \sum_{i,j;\alpha \neq \beta} x_i^\alpha x_j^\alpha x_i^\beta x_j^\beta \big].
\end{equation}

Thus,  Eq.(\ref{eq4.4}) can be further rewritten to have the  form
\begin{equation} \label{eq4.12}
\rho (\lambda) = \rho^{(I)} (\lambda) + \rho^{(M)} (\lambda) + \rho^{(K)} (\lambda)  
\end{equation}
 in terms of the and $I$,  $M$,  and $K$,  functions   defined here above. 

Further progress can be made in evaluating $I$, $M$,  and $K$,  to order unity such that 
$\rho^{(M)} (\lambda)$,  $\rho^{(K)} (\lambda)$, and  $ \rho^{(I)} (\lambda) $ be  known  to $ O(1/N)$, as follows:
\begin{itemize}
\item 
evaluating $M$ (Eq.(\ref{eqM}) 

One finds  to $O(1)$
\begin{equation}    \begin{array} {l}   \label{eq4.15}
M=  exp   \Big[nN \Big[   ln \Big( \frac{\pi}{i\lambda(1+s^\alpha)}  \Big)  \Big] ^{1/2}
 \;\\\\\ \ \ \ \    \ \ \ \    \ \ \ \  \ \ \ \    \ \ \ \  \ \ \ \  +Ê ln \Big[ 1 - i\frac{J^2}{N} p^\alpha  \frac{3}{(2i\lambda(1+s^\alpha))^2} \Big]
 \Big]
 . \end{array}  \end{equation}

When
 substituting the  $O(1)$ term  into Eq.(\ref{eqM}), assuming replica symmetry and using  the identity Eq.(\ref{eq2.3}) to reconstruct the logarithm, we obtain 

  \begin{equation}   
          \begin{array} {l}
\rho^{(M)} (\lambda)=   \frac{-2}{N\pi} Im \frac{\partial}{\partial\lambda}  ln\Big\{ \Big (\frac{e^{i\pi/4}}{\pi^{1/2}}\Big)^{N}   \Big(\frac{N}{2 \pi  }\Big)^{1/2}  \frac{\lambda}{(2J^2)^{1/2}}\\ \\\ \ \ \ \ \ \int ds\;   exp\Big[- \frac{\lambda^2}{4J^2}Ns^2+Nln(\frac{\pi}{i\lambda(1+s)})^{1/2} \Big]  \\ \\\ \ \ \ \ \ \   \int dp\;  (\frac{3}{8 \pi  })^{1/2}\;  exp\Big[  \frac{-3}{8}p^2 -iJ^2p \frac{3}{(2i\lambda(1+s))^2} \Big]
     \Big\}        \end{array}  \label{eq4.17}  . \end{equation}

The $p$-integral in Eq.(\ref{eq4.17}) contains only an exponential and can be easily evaluated.  Thus, $ \rho^{(M)} (\lambda)$  takes the concise form
  \begin{equation}   
          \begin{array} {l}
\rho^{(M)} (\lambda)=   \frac{-2}{N\pi} Im \frac{\partial}{\partial\lambda}  ln\Big\{    \lambda\; exp \Big(\frac{-N}{2}\;ln\lambda \Big ) \\ \\\ \ \ \ \ \ \ \ \ \ \ \int ds\;   exp[-Ng(s) +h(s)] 
     \Big\}    \end{array}  \label{eqrhoM}   \end{equation}
     
     where
       \begin{equation}   
       \label{eqrhogs}  
       g(s)=  \frac{\lambda^2 s^2}{4J^2 }  + \frac{1}{2}\; ln\Big[i\;(1+s)\Big]  \end{equation}
     and 
  \begin{equation}   
  h(s)= -\frac{3}{8}J^4\; \frac{1}{[i\lambda(1+s)]^4}
  \label{eqrhohs}   \end{equation}
 
\item 
A similar procedure goes for evaluating $K$ (Eq.(\ref{eqK})) 
  (see also DJ \cite{DJ90}).
Notice that there is no $p$ variable in $K$,  whence the $p$-integral  = 1; thus,  

  \begin{equation}   
          \begin{array} {l}
\rho^{(K)} (\lambda)=   \frac{-2}{N\pi} Im \frac{\partial}{\partial\lambda}  ln\Big\{    \lambda\; exp \Big(\frac{-N}{2}\;ln\lambda \Big  ) \\ \\\ \ \ \ \ \ \ \ \ \ \   \int ds\;   exp[-Ng(s) +f(s)] 
     \Big\}    \end{array}  \label{eqrhoK}   \end{equation}
     where
       \begin{equation}   
       \label{eqrhofs}  
       f(s)=  \frac{1}{4}\; ln\Big[1-\frac{J^2}{ \lambda^2(1+s) ^2}\Big]  \end{equation}

\item 
Evaluating $I$, (Eq.(\ref{eqI})), 
   is more straightforward; thereafter, one obtains
 
  \begin{equation}   
          \begin{array} {l}
\rho^{(I)} (\lambda)=   \frac{-2}{N\pi} Im \frac{\partial}{\partial\lambda}  ln\Big\{    \lambda\; exp\Big (\frac{-N}{2}\;ln\lambda  \Big) \\ \\\ \ \ \ \ \ \ \ \ \ \ \int ds\;   exp[-Ng(s) )] 
    \Big \}    \end{array}  \label{eqrhoI}   \end{equation}

\end{itemize}
The  above results can be easily collected to rewrite $ \rho (\lambda)$, Eq.(\ref{eq4.12}),
exactly to $O(1/N)$.

\section{AED and Green function to $O(1/N)$}\label{sec5} 
The method of steepest descent allows to obtain the AED to $O(1/N)$.
The interesting saddle point  $s_0$ associated to $g(s)$ is found from

\begin{equation} \label{eq5.16}
\frac{\partial g}{\partial s}\Big| _{s=s_0}=  0
\end{equation}

Thus,  \begin{equation}  \begin{array} {l} \label{eq5.2a}
\rho^{(M)} (\lambda)=    \frac{-2}{N\pi} Im \frac{\partial}{\partial\lambda}  \Big \{  -\frac{N}{2}  ln\lambda -Ng(s_0)  \\ \\\ \ \ \ \ \ \ \ \ \ \ +ln \lambda -\frac{1}{2}  ln\; g''(s_0) + h(s_0)\Big \} \end{array}.
\end{equation}
Similar expressions are obtained for $ \rho^{(N)} $ and  $\rho^{(I)}$, when replacing  $h(s_0)$ by $f(s_0)$  and 1  in Eq.(\ref{eq5.2a}), respectively.

In other words, to $O(1/N)$,  one has
 \begin{equation} \label{eq5.3}
 \rho (\lambda) =  \rho_0 (\lambda) +  \rho_{1/N} (\lambda) 
\end{equation}
with 
 \begin{equation} \label{eq5.3a}
 \rho_0 (\lambda) = \frac{-2}{N\pi} Im \frac{\partial}{\partial\lambda}    ln\Big\{-\frac{N}{2}ln\lambda -Ng(s_0)\Big\}
\end{equation}
 \begin{equation} \label{eq5.3b}
 \rho_{1/N} (\lambda) =  \frac{-2}{N\pi} Im \frac{\partial}{\partial\lambda}    ln\Big \{\lambda -\frac{1}{2}  ln\; g''(s_0) + h(s_0)+ f(s_0)\Big\}.
\end{equation}

 From the definition  Eq.(\ref{eqrhohs}), 
  it is easy to see that  $g  (s;\lambda)$ has two  complex conjugate saddle points at 
 \begin{center}
$ \frac {1}{2} \big[ -1\mp i (1-(4J^2/\lambda^2) ) \big]^{1/2} $. 
 \end{center}
It  has been argued in Edwards and Jones  \cite{EJ76}  that the contour chosen for the saddle point approximation may only be deformed to pass through one of these saddle points, and  that the lower saddle point leads to a physically reasonable positive AED. Thus, following Edward and Jones   \cite{EJ76},  we choose  the $-i$ sign saddle point, in the above expression, to go on.


We now explicitly evaluate $ \rho (\lambda)$. The contribution $ \rho_0 (\lambda)$, obtained from Eq.(\ref{eq5.3a}), taking into account  Eq.(\ref{eqrhogs}),  
is

 \begin{equation} \label{eq5.5}
   \rho_0 (\lambda) = \frac{1}{\pi}\;Im\Big\{\frac{1}{2J^2}\Big[\lambda+i\;(4J^2-\lambda^2)^{1/2}\Big]\Big\}
\end{equation}

yielding the corresponding Green function
 \begin{equation} \label{eq5.6}
G_0 (\lambda) = \frac{1}{2J^2}\;\Big[\lambda+i\;(4J^2-\lambda^2)^{1/2}\Big]
\end{equation}
thereby proving that

 \begin{equation} \label{eq5.7}
   \rho_0 (\lambda) =\left\{ \begin{array}{lcl}
(\frac{1}{2\pi J^2} ) \;[4J^2-\lambda^2 ]^{1/2},  \ \   \mbox{\ \ for \  } |\lambda |<2J\\ 
 \ \  \ \   \ \  \ \  \ \  \ \  0, \ \  \ \  \ \  \ \  \ \    \ \ \  \mbox{\ \ \  \ for \ \ \  }  |\lambda |>2J.%
\end{array}%
\right.  \end{equation}
 
 The first order correction $ \rho_{1/N} (\lambda) $ is obtained after some simple algebra, taking into account Eq.(\ref{eqrhohs})  and  Eq.(\ref{eqrhofs}), reads
  \begin{equation} \label{eq5.8}
 \rho_{1/N} (\lambda) =\left\{ \begin{array}{lcl} \frac{1}{4N}\;\Big[\delta(\lambda+2J) +\delta(\lambda-2J) \Big] -  \frac{1}{2N\pi}   \frac{1}{[4J^2-\lambda^2 ]^{1/2}}\;  \\ \\  \ \ + \frac{3}{ N\pi}  \frac{  [4J^2-\lambda^2 ]^{1/2}}{8J^4}\Big \{ [3\lambda^2-2J^2 ] -\frac{2\lambda^2(\lambda^2-2J^2)}{[4J^2-\lambda^2 ] }\Big  \},  \\ \\  \ \ \ \  \ \  \ \  \ \  \ \  \ \  \ \  \ \  \ \  \ \   \ \ \ \ \ \ \mbox{\ \ for \  } |\lambda |<2J\\  \\
 \ \  \ \   \ \  \ \  \ \  \ \  0, \ \  \ \  \ \  \ \  \ \  \ \   \ \ \  \mbox{\ \ for \ \  }  |\lambda |>2J.%
\end{array}%
\right.  \end{equation}

Thereafter, the first order Green function correction $G_1(\lambda)$ is immediately obtained  from $ \rho_{1/N} (\lambda)$. Moreover,  the former  and the latter functions can be expressed in terms of the zero order Green function or AED respectively. After some lengthy, but simple,  algebra, it is found that

\begin{equation}\label{eq5.11}
G  (\lambda) = G_0  \Big\{1+ \frac{1}{N}\Big[\frac{J^2G_0^2}{(1-J^2G_0^2)^2}-3\frac{(JG_0)^4}{1-J^2G_0^2} \Big] \Big\}
\end{equation}
where the variable $(\lambda) $ has not be written in the r.h.s, and $G_0$ is defined in Eq.(\ref{eq5.6}).

 Eqs.(\ref{eq5.7})-(\ref{eq5.11}) are the new intended results, whence presenting the extra terms not found in previously treated cases, e.g. \cite{DJ90}, arising from the symmetry  and sign of the element constraint imposed on the Wigner matrix.

\section{Comments}
\label{comments}
For a short discussion, let us decompose  $ \rho_{1/N} (\lambda) $ such that

  \begin{equation} \label{eq5.12}
 \rho_{1/N} (\lambda) =  \rho_{1/N}^{(Q)} (\lambda) +  \rho_{1/N}^{(R)} (\lambda)
 \end{equation}
 
thus,  where 
 \begin{equation}       
 \rho_{1/N}^{(Q)} (\lambda)  =     \frac{1}{4N}\;[\delta(\lambda+2J) +\delta(\lambda-2J) ] 
 - \Big\{ \frac{1}{2N\pi}   \frac{1}{[4J^2-\lambda^2 ]^{1/2}}\Big\} \;,
 \label{eq5.12a} \end{equation}
which   is identical to Eq.(4.14) of DJ  \cite{DJ90} and
\begin{equation}         \begin{array} {l} \label{eq5.12b}
 \rho_{1/N}^{(R)} (\lambda) =   \frac{3}{ N\pi}  \frac{  [4J^2-\lambda^2 ]^{1/2}}{8J^4} \Big\{ [3\lambda^2-2J^2 ] -\frac{2\lambda^2(\lambda^2-2J^2)}{[4J^2-\lambda^2 ] } \Big\}\; ,
\end{array}\end{equation}
This Eq.(\ref{eq5.12b}) is the extra  term coming from the Feynman diagrams  presented here above, which added to the Eq.(\ref{eq5.12}) corresponding to  the $O(1/N)$ correction to the Gaussian Orthogonal Ensemble distribution, is one key point of our present work.
Here, it becomes clear that the  first correction for   the RSSME arises from $ \rho_{1/N} (\lambda)$,  which in turn is based on the function $h(s)$ defined by Eq.(\ref{eqrhohs}).

On  Figs. \ref{fig: ScFig1NxN2x2}-\ref{fig: ScFig5NxN200x200}, we display $ \rho_0 (\lambda)$ and $ \rho  (\lambda)$ for $N$ = 2, 10, 20, and 200 respectively. For convenience $J$ is taken to be  =1. We notice that there is a significant departure of $ \rho  (\lambda)$ from the semicircle $ \rho_0 (\lambda)$ even for values of $N$ as large as 20.  For the GOE,  the departure from the semicircle becomes noticeable only for N $<$ 6  \cite{VZ, JD}.   By comparison with the figures supplied in DJ \cite{DJ90},  it can be seen that the significant departure is due to  $\rho_{1/N} ^{(R)} (\lambda)$   rather than  $\rho_{1/N}^{(Q)} (\lambda)$.  From this we deduce that it is this new correction   $ \rho_{1/N}^{(R)} (\lambda)$    together with $ \rho_0 (\lambda)$ which mimics  the broadening   of the two mirror imaged Poisson type distributions as $N$ becomes large. In the limit of large $N$ ($\sim  $   200), this broadening and overlapping tend  to the semicircle. One should not expect  $ \rho (\lambda)$  to mimic correctly the AED of the RSSME  when $ N <$ 6.  Corrections  to $O(1/ N^2)$ will be required for these low values of $1/N$. They will also be required when mimicking the fine structure of the spectrum. 

The displayed  figures also bring to the fore that  $\rho_{1/N} (\lambda)$   possesses divergences near the band edges $  |\lambda |=2J$ of the semicircle.  This is not surprising for  reasons mentioned earlier. Thus, result $ \rho_0 (\lambda)$, combining Eq.(\ref{eq5.7}) and Eq.(\ref{eq5.8}), should only be  considered as  best away and inside the band edges. 

Briefly,  we finally comment on the result in Eq.(\ref{eq5.11}). Similarly we decompose $ G_1(\lambda)$, as done for   $\rho_{1/N} (\lambda) $, into 

  \begin{equation} \label{eq5.13}
G_1 (\lambda) =  G_1^{(Q)} (\lambda) + G_1^{(R)} (\lambda)
 \end{equation}
 whence with 
  \begin{equation} \label{eq5.13a}
G_1^{(Q)} (\lambda) =  \frac{1}{N}\Big[\frac{(JG_0)G_0(JG_0)}{(1-(JG_0)^2)^2} \Big]
 \end{equation}
 and
   \begin{equation} \label{eq5.13b}
G_1^{(R)} (\lambda) =  - \frac{3}{N}\Big[\frac{(JG_0)^2G_0(JG_0)^2}{(1-(JG_0)^2)}  \Big].
 \end{equation}
 
 It can be noticed that  $G_1^{(Q)} (\lambda) $ is the first order correction of the GOE, while 
 $ G_1^{(R)} (\lambda) $ is  the newly found  first order correction to the RSSME.

 \section{Numerical Simulations}\label{sec6NumericalSimulation} 
 
Numerical simulations have proceeded as follows: first,  the matrix size $N$ is decided upon, and  zeroes are put on the diagonal. Next one picks at random an element, $a_{i,j}$ of the matrix, with  $1\le i \le N$ and $1\le j \le N$; one attributes  either the value $+J/\sqrt  N$ or  $-J/\sqrt N$ with equal probability to  $a_{i,j}$ AND to $a_{j,i}$. (One can take $J=1$ without loss of generality). Do such an  attribution for another $a_{i,j}$ element, in fact  successively for all  $N(N-1)$  elements of the matrix. Calculate the $N$ eigenvalues, and store them. Repeat the matrix construction a large number of times.    In the present case, all $N$-size matrices were  invented a million of times, except for $N=200$     only 50000 times.  

The histogram of eigenvalues for the set of $N$-given finite size matrix is thus obtained. Practically, the  histogram is  normalized according to the number of simulations, in order to obtain the "averaged spectrum". 
The display of such AED is shown in Fig. \ref{fig: SimFig1NxN2x2} and Figs. \ref{fig: SimFig4NxN10x10}-\ref{fig: SimFig11NxN200x200}. On such figures, the theoretically obtained first order finite size correction   $\rho(\lambda)$ is also given for comparison.
 

Noticed that each numerical spectrum seems to have some nice band tails. 

 \section{Conclusions}\label{sec7conclusions} 

Using the  replica method, we have  searched for the first finite size  $O(1/N)$ correction to the Averaged Eigenvalue Density  $and$ to the  corresponding Green function of a    random sign symmetric matrix ensemble.   It is well known that  the AED of a $d$-regular random graph with $N$ vertices, in the limit  $N  \rightarrow \infty$   and $d  \rightarrow \infty$}, obeys the Kesten-McKay law \cite{KMlawKesten,KMlawMcKay}. However, fully random systems are only theoretical cases of  interest. Thus it seemed worthwhile to calculate correction terms to the AED in view of handling more realistic systems. In  our work, the former  correction term  to $O(1/N)$  becomes written as in Eq.(\ref{eq5.8}), while the total Green function correction term reads as in Eq.(\ref{eq5.11}):

\begin{equation}\label{eq5.11b}
G_1  (\lambda) = G_0 \; \frac{1}{N}\Big[\frac{J^2G_0^2}{(1-J^2G_0^2)^2}-3\frac{(JG_0)^4}{1-J^2G_0^2} \Big]  
\end{equation}
 
The interpretation of the extra terms seems rather clear, pertaining to the reduced  number of  "degrees of freedom"  of the system, within the Hamiltonian and the corresponding matrix of   (a reduced number of) possible states; the "restriction" being found  in the  equal probability condition for the $binary$ distribution of matrix elements, but in the "extension"  in the  $(\pm)$ sign of these matrix elements.  

    It can be usefully reinstated that  the term in brackets in Eq.(\ref{eq3.4})  leads to the relevant term in Eq.(\ref{eq4.4}), thereby allowing to  distinguish  the difference
between
matrices
with
Gaussian
distributed
matrix
elements
and
those
with
the
binary
distribution
considered
in this paper. This term  subsequently sustains Eq.(\ref {eq4.15}  ) and Eq.(\ref{eqrhohs})  for  the  (1/N) correction to $\rho(\lambda)$, whence going beyond DJ analysis   \cite{DJ90}.

Beside analytic works,  simulation  results have been presented. The comparison between the analytical formulae and the numerical diagonalization results for  finite size matrices  exhibits an excellent agreement, confirming the correctness of  the first order finite size expression.

It has been emphasized  that the $1/N$ corrections of the AED  diverge at the band edges of the semicircle.  This ''problem''  should be considered in  further work. Some   self consistency condition imposed on the diagrammatic formalism should produce  a finite AED throughout the whole spectrum.  However, this is obviously outside the present aim. 

More recently,  studies of the properties of random matrices have found a new revitalisation due to the mapping of networks and graphs through their adjacency matrix. In these cases, an additional input stems from the possible directionality of the link or bond. Let us have in mind, for argument, the case of a citation or any type of cooperation/competition network. Due to the  intrinsically time dependent hierarchical  process,  the adjacency matrix representing the network is usually asymmetric, beside being a non-negative matrix.    This asymmetry, leading to complex eigenvalues, much widens the realm of investigations \cite{physaGR,GRMAEPJB86}. The asymmetry case is \underline{not} treated in the present paper, but is mentioned  in this conclusion section, for any  reader guideline   interested  in   pursuing the present work.

Thus, further work, if we may suggest so, should be  programmed in order to apply the approach in view of  obtaining results  on disordered systems characterized, e.g., (i) by symmetric matrices  having more complicated structures, as in financial or socio-economic matter  \cite{gligorausloosAPPA,physaGR,GRMAEPJB86}), or  (ii) when more than one type of disorder  appears, as is the case very often  in materials \cite{PRB4FAA,MapleFischer82,ESR100.10.1 mineralogy}, 
and (iii) by non-symmetric matrices: see the  cases of citation or coauthorship networks  implying link ordering\cite{scim101.14.587coauthors,PRE91.15.012825coauthor,PRE80.09.046110rdmgraph}, that of bipartite graphs \cite{1JSM}, that of physiology \cite{Kwapienetalbrain}, or that of financial markets \cite{plerou1999universal,utsugi2004random,Kwapienfinancial,Lucaasymmcorr,bielythurner}, among recent relevant cases.

  \begin{figure}
\includegraphics[height=6.0cm,width=8.4cm] {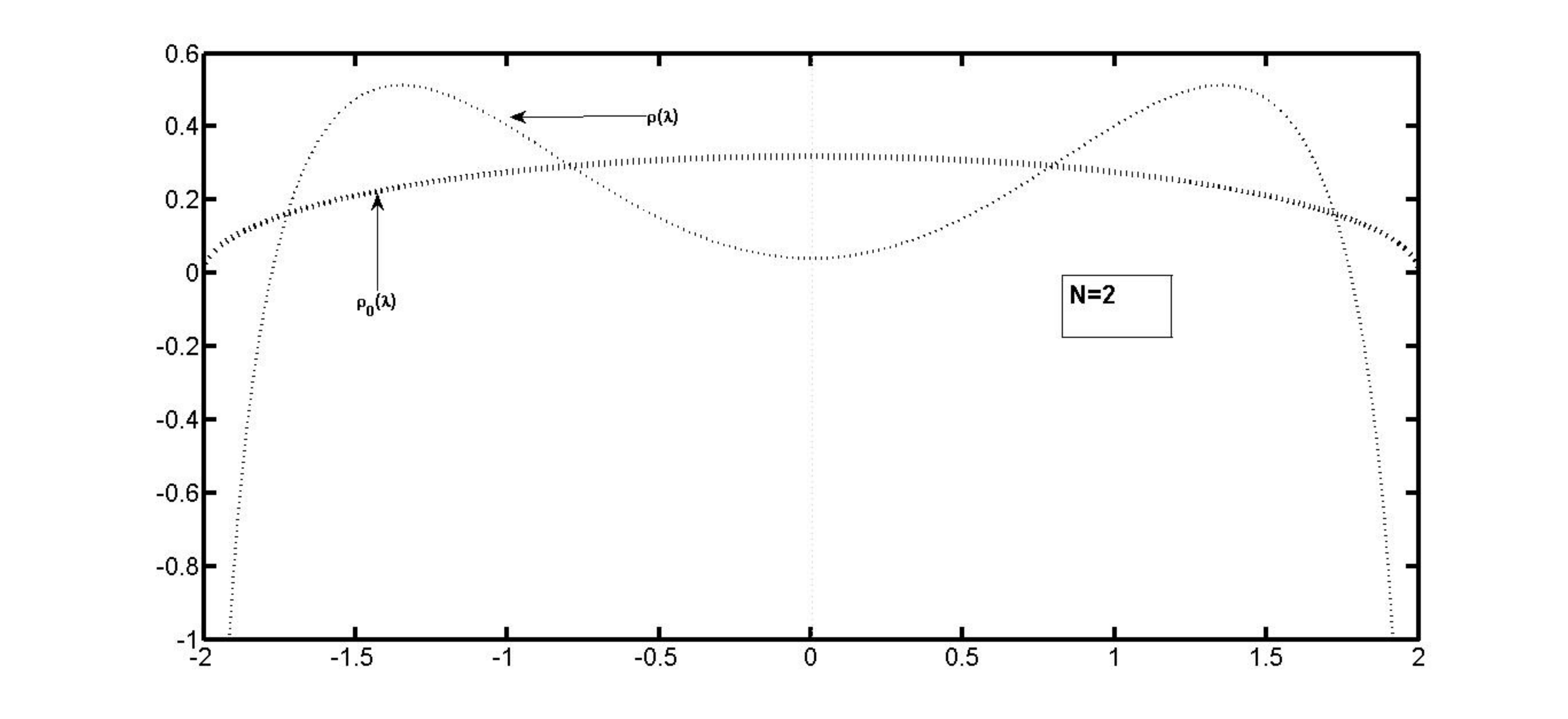}
  \caption   {
  Plot of  $\rho_0(\lambda)$  and $\rho(\lambda)$  for $N =2$.  }
 \label{fig: ScFig1NxN2x2}
\end{figure}


  \begin{figure}
\includegraphics[height=6.0cm,width=8.4cm] {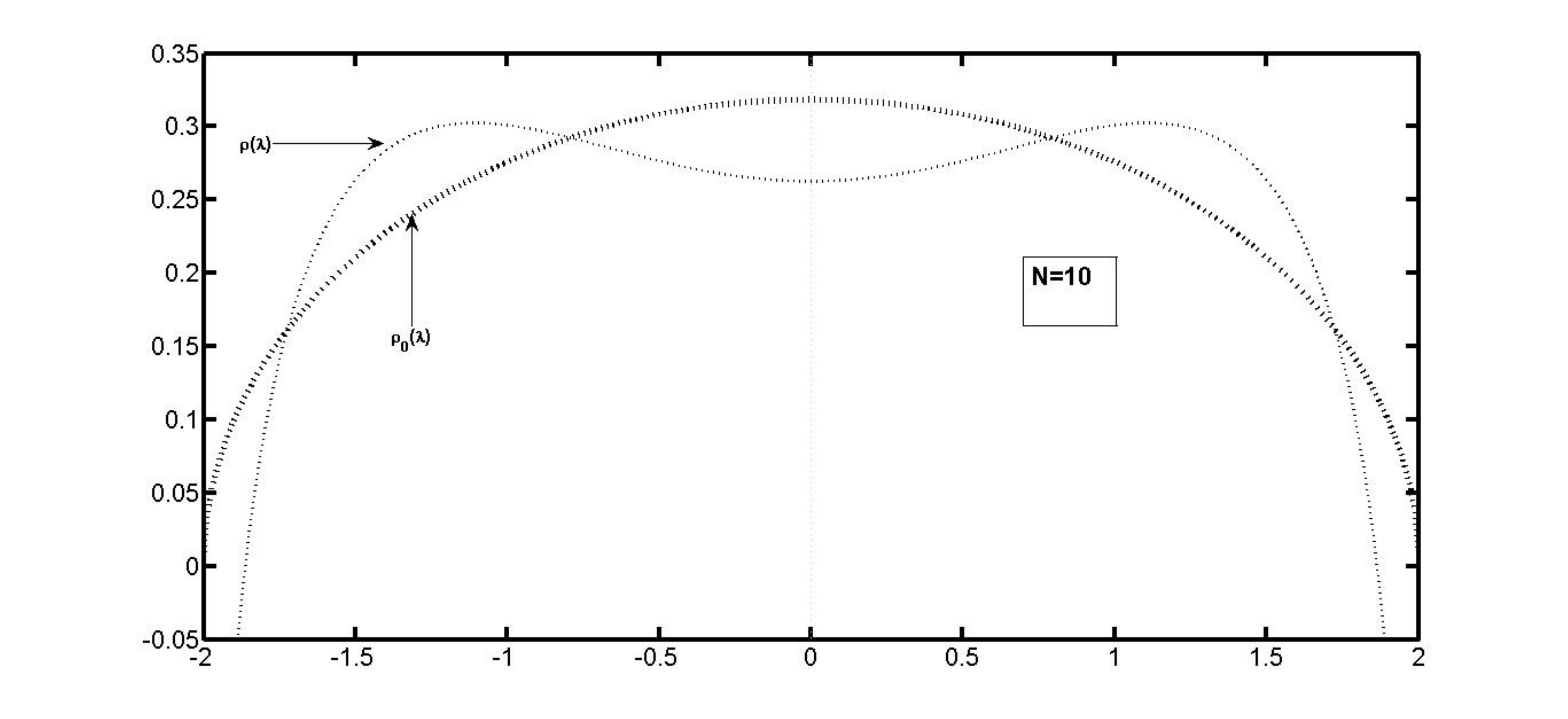}
  \caption   {
 Plot of  $\rho_0(\lambda)$   and $\rho(\lambda)$   for $N =10$.  }
 \label{fig: ScFig3NxN10x10}
\end{figure}

  \begin{figure}
\includegraphics[height=6.0cm,width=8.4cm] {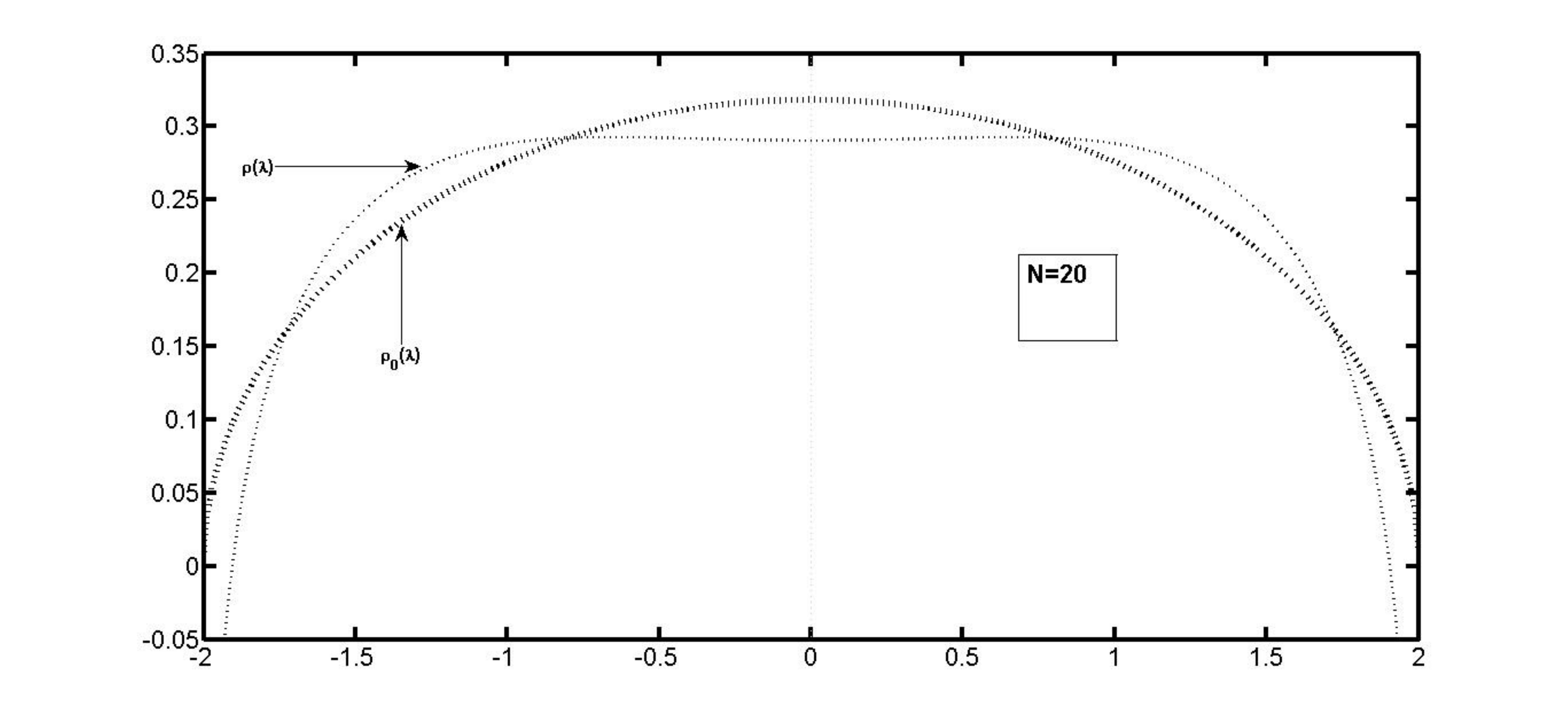}
  \caption   {
 Plot of  $\rho_0(\lambda)$   and $\rho(\lambda)$     for $N =20$.  }
 \label{fig: ScFig4NxN20x20}
\end{figure}

  \begin{figure}
\includegraphics[height=6.0cm,width=8.4cm] {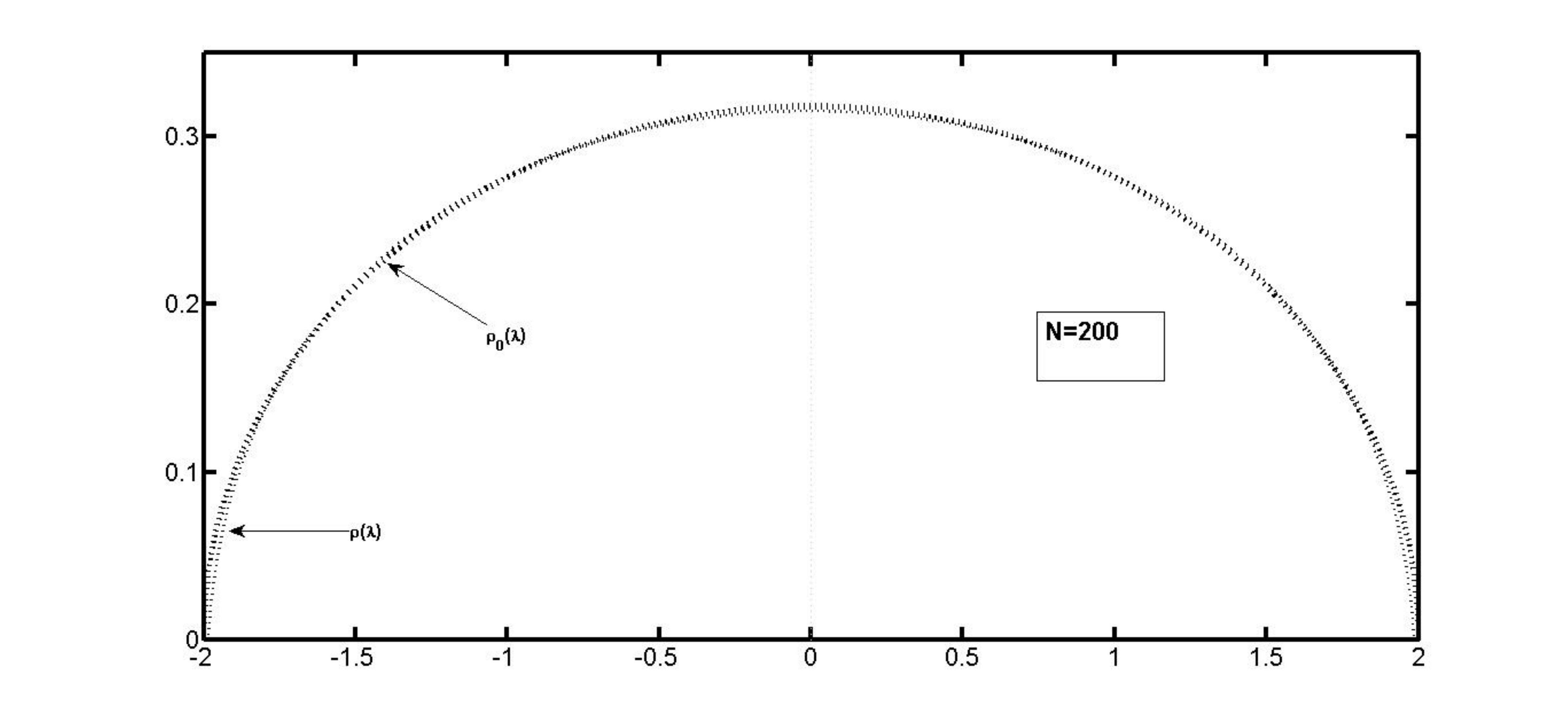}
  \caption   {
  Plot of  $\rho_0(\lambda)$    and $\rho(\lambda)$  for $N =200$.  The curves are hardly distinguishable.}
 \label{fig: ScFig5NxN200x200}
\end{figure}
  \vskip0.3cm

  \begin{figure}
\includegraphics[height=6.0cm,width=8.4cm]{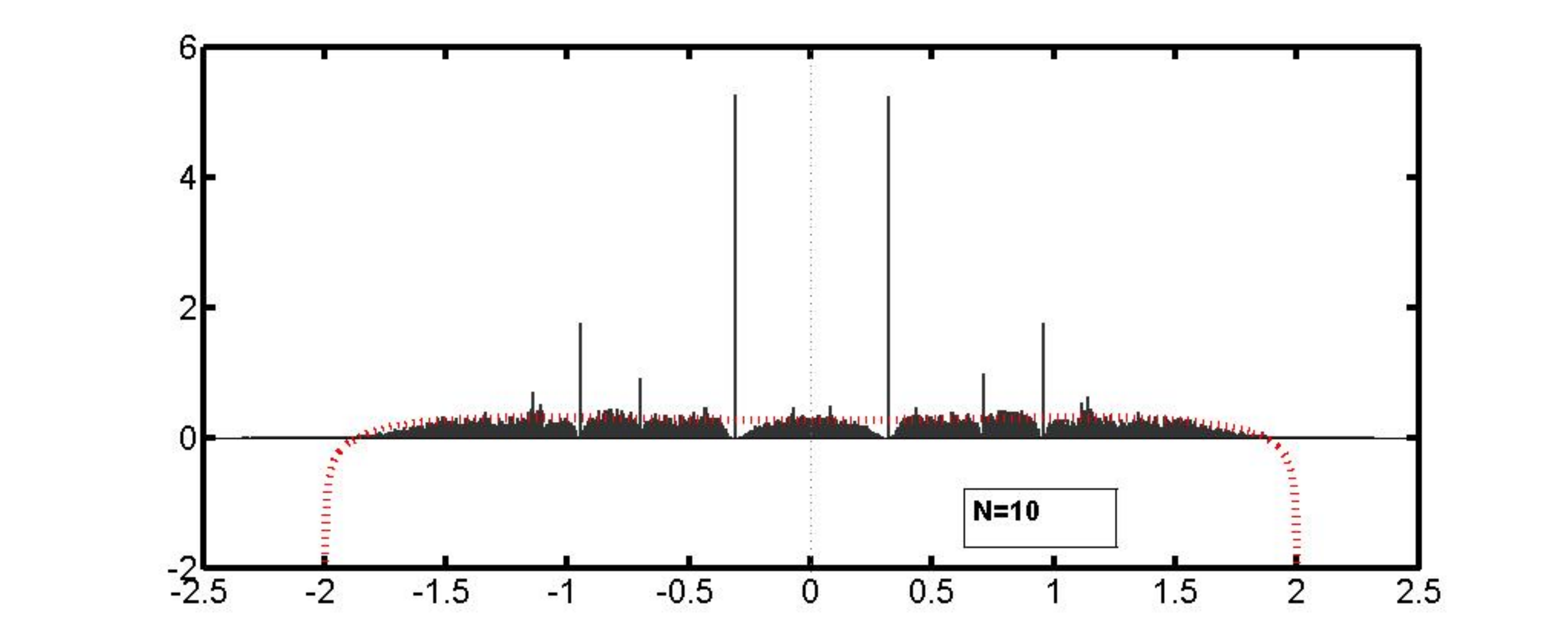}
  \caption   {
 Comparison of the numerical simulation AED  (vertical lines, delta functions) with the theoretical first order $O(1/N)$ approximation (red) dotted line 
 for $N=10$.   }
 \label{fig: SimFig4NxN10x10}
\end{figure}

  \begin{figure}
\includegraphics[height=6.0cm,width=8.4cm] {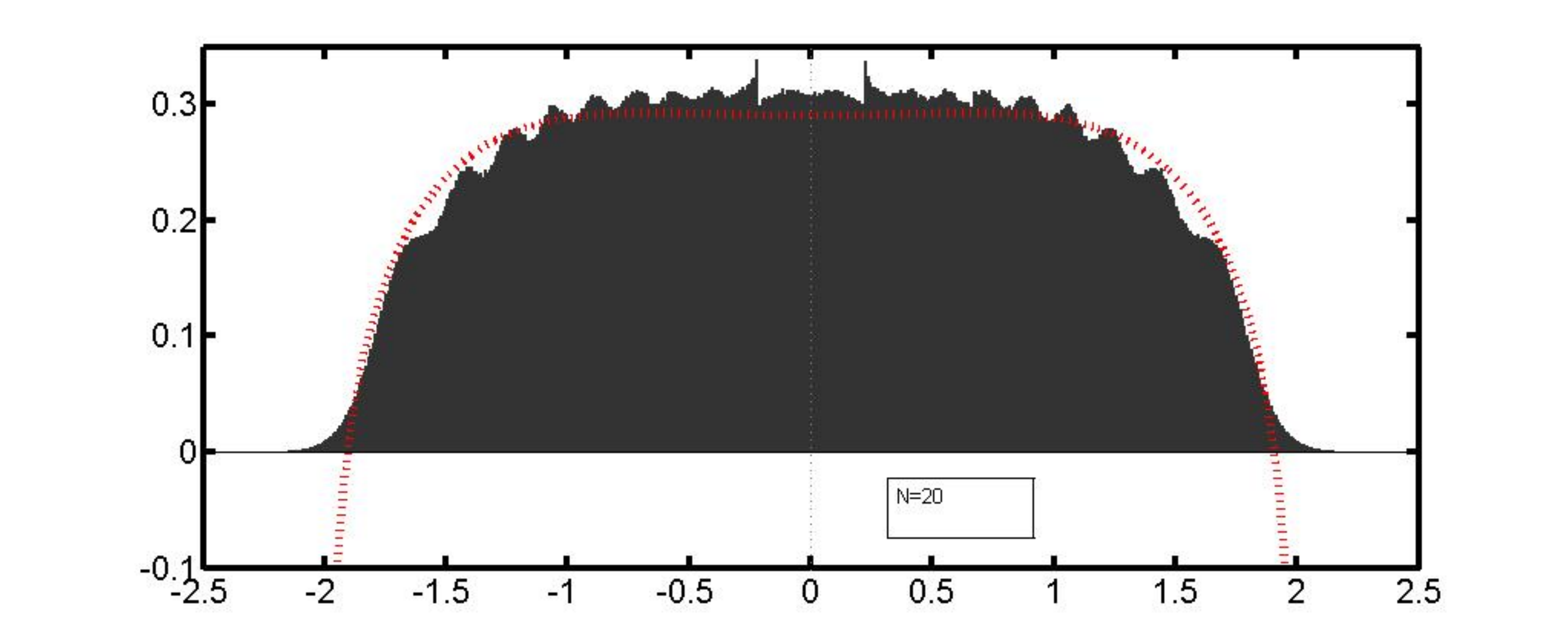}
  \caption   {
 Comparison of the numerical simulation AED  (vertical lines, delta functions) with the theoretical first order $O(1/N)$ approximation (red) dotted line 
  for $N =20$.   }
 \label{fig: SimFig7NxN20x20}
\end{figure}

  \begin{figure}
\includegraphics[height=6.0cm,width=8.4cm] {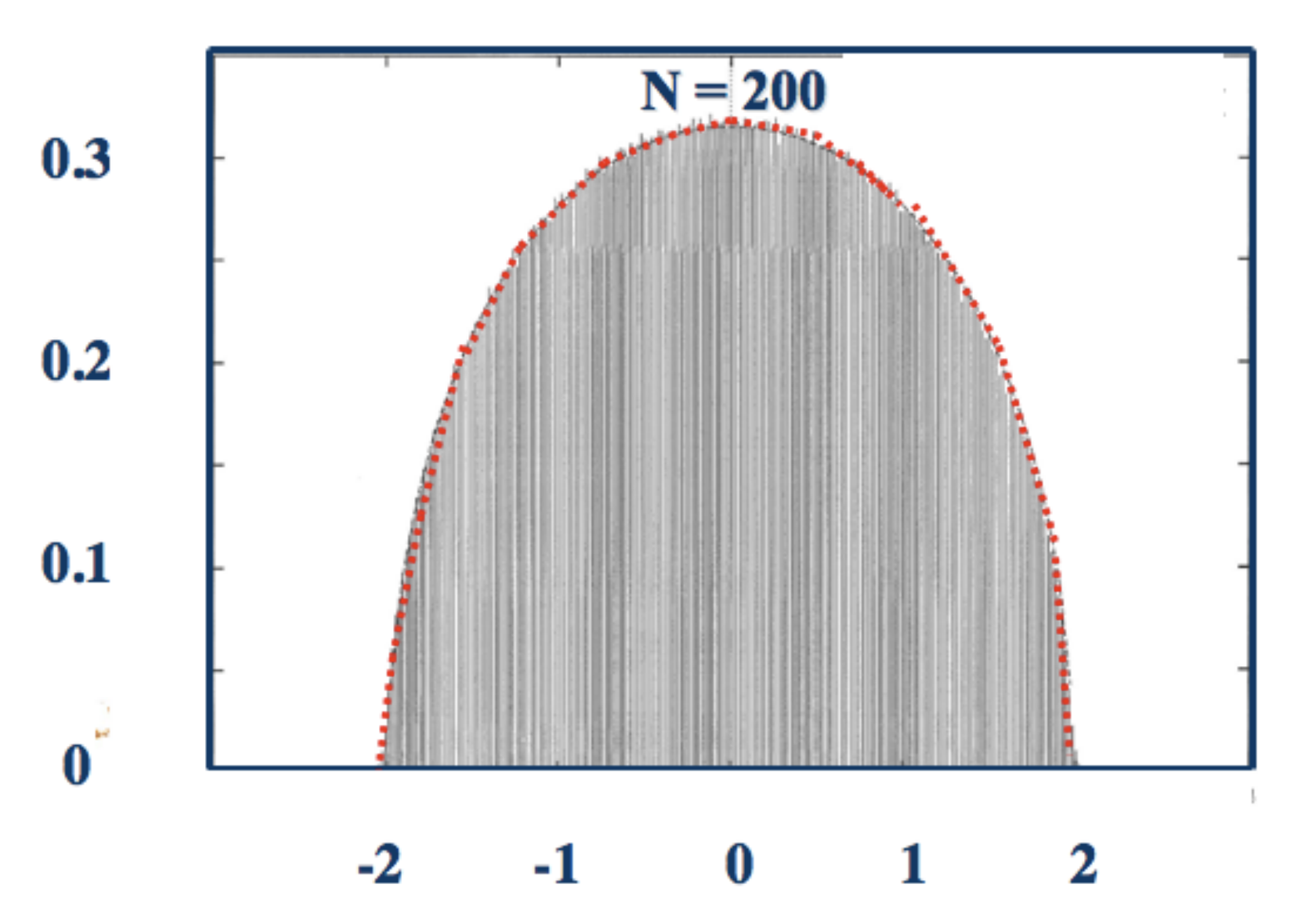}
  \caption   {
  Comparison of the numerical simulation AED  (blue) vertical lines, delta functions, with the theoretical first order $O(1/N)$ approximation (red) dotted line 
  for $N =200$. }
 \label{fig: SimFig11NxN200x200}
\end{figure}

{\bf Acknowledgements}  
 
 Great thanks to J. Miskiewicz  and M.B. Shakeel for much "technical help", and to anonymous reviewers for their comments.
 
\vskip0.4cm
{\bf Appendix. \\ \\On "neglecting" diagonal terms in the rhs of Eq.(13) } \vskip0.4cm
  Recall   Eq.(13) transforming a product of $cos$ into a product of $exp$,  apparently  including diagonal ($i=j$)
terms on the  rhs, but  not including them on the rhs.   Although this should be much incorrect indeed on  rigorous grounds, let it be recalled that 
  \begin{equation}
 \label{eq2.38}
 ln[cos\frac{A}{\sqrt(N)}] \simeq \; ln[1-\frac{A^2}{2N} + \dots] \simeq \;  -\frac{A^2}{2N} + O(\frac{1}{N^2})
 \end{equation}
to leading order in $N$. Therefore,

    \begin{equation}   
          \begin{array} {l}
  \rho(\lambda)  =-\frac{2}{N\pi} Im \frac{\partial}{\partial\lambda}  \lim_{n\rightarrow 0} \frac{1}{n}  \Big \{\Big(\frac{e^{i\pi/4}}{\pi^{1/2}}\Big)^{Nn} \; \\ \\\ \ \ \ \ \ \int_{-\infty}^{\infty}\prod_{i;\alpha}dx_{i;\alpha} \Big[ exp(-i\;\lambda \sum_{i;\alpha}(x_i^{\alpha})^2\Big]\; \\ \\\ \ \ \ \ \ \ \;\prod_{i<j}\Big[ cos(  \frac{2J}{\sqrt N}  \sum_{\alpha}x_i^{\alpha}x_j^{\alpha})-1\Big]\Big\}
             \end{array}    \end{equation}   \label{eqA1} 
             
             can be rewritten as in Eq. (14)
         \begin{equation}   
          \begin{array} {l}
                   \prod_{i<j}cos\Big(  \frac{2J}{\sqrt N}  \sum_{\alpha}x_i^{\alpha}x_j^{\alpha}\Big) \simeq
                   \;\\\  \\\ \ \ \   exp\Big\{ \sum_{i,j}\Big (  \frac{-J^2}{ N} ( \sum_{\alpha}x_i^{\alpha}x_j^{\alpha})^2\Big) \Big [1+\frac{2J^2}{3N}  ( \sum_{\alpha}x_i^{\alpha}x_j^{\alpha})^2\Big] \; \\ \\\ \ \ \ \    \ \ \ \    \ \ \ \  \ \ \ \    \ \ \ \  \ \ \ \    \ \ \ \    \ \ \ \  \ \ \ \    \ \ \ \     + O(N^{-3})\Big\}
             \end{array}    \end{equation} \label{eqA2}
 when neglecting the contribution of the diagonal terms. These read
               \begin{equation}   
          \begin{array} {l}
        \sum_{i,j}\Big (  \frac{-J^2}{ N} ( \sum_{\alpha} (x_i^{\alpha} )^2 )^2\Big) \Big [1+\frac{2J^2}{3N}  ( \sum_{\alpha}(x_i^{\alpha} )^2)^2\Big]  
           \end{array}    \end{equation} \label{eqA3}
           The first term
           \begin{equation}
       \simeq    \sum_{i}\Big (   ( \sum_{\alpha} (x_i^{\alpha} )^2 )^2\Big)               \end{equation}
              and the second term (coming for the consideration of finite size effects)
                  \begin{equation}
       \simeq   \sum_{i}\Big (   ( \sum_{\alpha} (x_i^{\alpha} )^2 )^4\Big)               \end{equation}
              give a contribution O($n^2$) and O($n^4$),  respectively, in $\lim_{n\rightarrow 0} $. Thus the diagonal elements in Eq.(14),  after the transformation resulting from  the Eq.(13)  expansion,  can be neglected in $\lim_{n\rightarrow 0} $.

             Notice that if we had started with the ensemble described by Wigner, i.e.  diagonal elements equal to zero,  we should have arrived to  Eq.(61) for  $  \rho(\lambda)$ as well. In this sense,  it is even irrelevant whether the diagonal elements  are  equal or not to zero.

  \end{document}